\documentclass[conference]{IEEEtran}
\IEEEoverridecommandlockouts
\usepackage{cite}
\usepackage{amsmath,amssymb,amsfonts}
\usepackage{algorithmic}
\usepackage{algorithm2e}
\usepackage{graphicx}
\usepackage{textcomp}
\usepackage{xcolor}
\usepackage{subfig}
\usepackage{hyperref}

\def\BibTeX{{\rm B\kern-.05em{\sc i\kern-.025em b}\kern-.08em
    T\kern-.1667em\lower.7ex\hbox{E}\kern-.125emX}}
\begin{document}

\title{Deep Attention-Based Alignment Network for Melody Generation from Incomplete Lyrics}


\author{
\IEEEauthorblockN{Gurunath Reddy M\IEEEauthorrefmark{1}\textsuperscript{$\dagger$	}\textsuperscript{$\ddagger$},
Zhe Zhang\IEEEauthorrefmark{1}\textsuperscript{$\dagger$},
Yi Yu\IEEEauthorrefmark{1},
Florian Harscoet\IEEEauthorrefmark{1}\textsuperscript{$\ddagger$}, 
Simon Canales\IEEEauthorrefmark{1}\textsuperscript{$\ddagger$} and
Suhua Tang\IEEEauthorrefmark{4}}
\IEEEauthorblockA{\IEEEauthorrefmark{1}Digital Content and Media Sciences Research Division\\
National Institute of Informatics, SOKENDAI}
\IEEEauthorblockA{yiyu@nii.ac.jp}
\IEEEauthorblockA{\IEEEauthorrefmark{4}Department of Computer and Network Engineering, Graduate School of Informatics and Engineering\\
The University of Electro-Communications}

}

\maketitle

\begingroup\renewcommand\thefootnote{$\dagger$}
\footnotetext{Equal contribution.}
\endgroup

\begingroup\renewcommand\thefootnote{$\ddagger$}
\footnotetext{Gurunath, Florian, and Simon were involved in this work during their internship in National Institute of Informatics (NII), Tokyo.}
\endgroup

\begin{abstract}
We propose a deep attention-based alignment network, which aims to automatically predict lyrics and melody with given incomplete lyrics as input in a way similar to the music creation of humans. Most importantly, a deep neural lyrics-to-melody net is trained in an encoder-decoder way to predict possible pairs of lyrics-melody when given incomplete lyrics (few keywords). The attention mechanism is exploited to align the predicted lyrics with the melody during the lyrics-to-melody generation. The qualitative and quantitative evaluation metrics reveal that the proposed method is indeed capable of generating proper lyrics and corresponding melody for composing new songs given a piece of incomplete seed lyrics.
\end{abstract}

\section{Introduction}

Automatic music generation aims to create meaningful music based on machine-assisted environments, which has been a challenging research issue in artificial intelligence and music. In the past few years, we have witnessed an increasing amount of research work on automatic music generation. With the advent of exploded music collections available on the Internet, data-driven methods such as Hidden Markov models~\cite{pachet2011markov}, graphic models~\cite{pachet2017sampling}, and deep learning models~\cite{waite2016generating,chu2016song,mogren2016c,dong2018musegan,yu2019deep}, have shown a potential for music creation. There exists a substantial amount of research on unconditional music generation, while less work considers generating melody from lyrics when given complete or incomplete lyrics. The primary reasons for substantially less research on conditional melody generation can be attributed to i) the non-availability of lyrics-melody paired dataset, ii) a lyrics composition can have multiple corresponding melodic representations, which makes it hard to learn the correlation between the lyrics and melodies, and iii) it is hard to evaluate the generated melodies objectively.

This work focuses on the most challenging aspect of music generation which enables deep learning models to discover proper lyrics and meaningful melodies when given a few words as input. To the best of our knowledge, this is the first attempt to generate melodies from incomplete lyrics, which is applicable to those who have no music knowledge but want to compose a song. Specifically, we present the lyrics-to-vector (lyric2vec) model which is trained on a large dataset of popular English songs to obtain the dense representation of lyrics at syllables, words, and sentence levels. The proposed lyrics-to-melody net (LTMN) is trained based on deep attention-based alignment in an encoder-decoder way. Moreover, we prove the importance of dense representation of lyrics by various qualitative and quantitative measures. Extensive experiments have demonstrated the effectiveness of the proposed LTMN which can automatically generate proper lyrics and meaningful melody when given a few words as seed lyrics.


\section{Related Work}
\label{sec:rel_work}

Lyrics-conditional music generation task is explored by researchers in the past few years. ALYSIA~\cite{ackerman2017algorithmic} generates melodies from the lyrics based on random forests. An encoder-decoder RNN sequential architecture for lyrics-conditional melody generation for Chinese pop songs is presented in~\cite{bao2018neural}. A large-scale paired lyrics-melody dataset with alignment information is created and the generation of melody from given lyrics of full length is studied in~\cite{yu2019conditional}. The authors investigate an end-to-end deep melody generative network conditioned on lyrics, where conditional long short-term memory - generative adversarial network (LSTM-GAN) was proposed for melody generation from lyrics. The authors in~\cite{yu2020lyrics} demonstrated the effectiveness of the proposed melody generation system by generating pleasing and meaningful melodies matching the given lyrics. Three Branch Conditional LSTM-GAN was proposed to improve the generation quality with Gumbel-Softmax technique in~\cite{srivastava_melody_2022}. In addition, the authors in~\cite{ShengST2021} extended a pre-trained model to better capture contextual information. An interpretable lyrics-to-melody generation system was proposed in \cite{duan_interpretable_2022}. A conditional hybrid GAN model was proposed for melody generation from lyrics in \cite{yu_conditional_2022}.

Distinguished from the existing methods, this work addresses how to generate proper lyrics and meaningful melodies when given incomplete lyrics. Particularly, meaningful lyrics are first generated from the given seed lyrics with LSTM networks. Taking each syllable of lyrics as input, the sequences of music attributes (pitch, duration, and rest) are predicted accordingly. The attention technique is proposed to model the sequence for lyrics-to-melody generation.

\begin{figure}[htbp]
        \centering
        \includegraphics[width=0.45\textwidth]{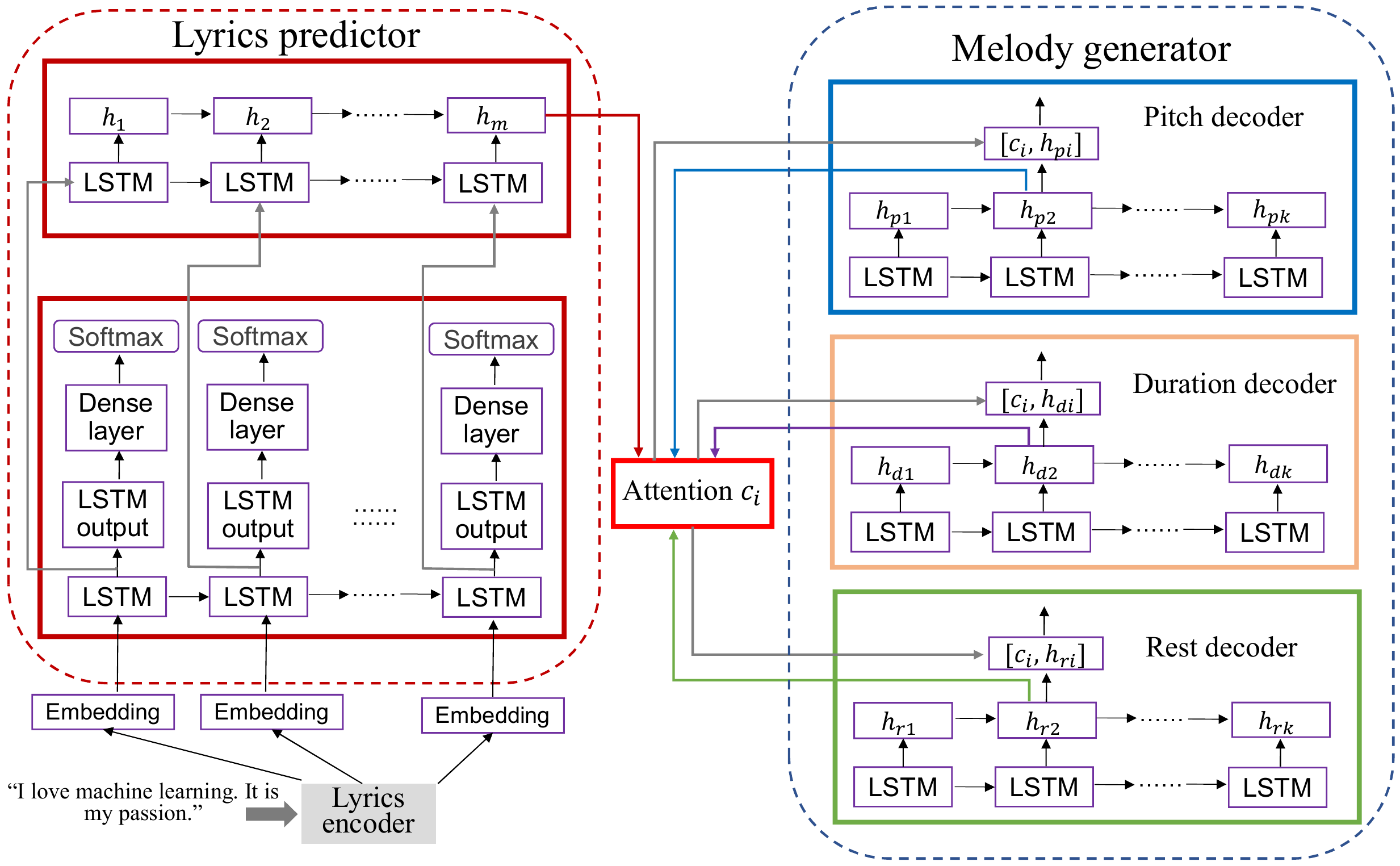}
        \caption{Our LTMN mainly consists of a lyrics prediction module and a melody generation module. Lyrics are predicted by an LSTM-based lyrics predictor. Melodies are predicted by individual decoders of pitch, duration, and rest.}
        \label{fig:neural_lyrics_melody}
 \end{figure}
 
\section{Proposed Lyrics-to-Melody Generation}
\label{sec:prob_def}
 
In our work, an attention-based network is designed to generate melodies from incomplete lyrics. The model first generates lyrics given seed lyrics and then composes the melody for the generated lyrics . We can model our music generation as a probabilistic model with two conditionally dependent components: a lyrics predictor and a lyrics-to-melody generator as shown in Fig.~\ref{fig:neural_lyrics_melody}.

The lyrics predictor is modeled as a conditional distribution to predict the next lyrical token given the previous tokens of the lyric sequence modeled as $p(s_t | s_1, s_2, ..., s_{t-1})$. Here, the tokens are the sequence of syllables of the lyrics given by $S = \{s_1, s_2, . . ., s_T\}$. We can learn a probability distribution
\begin{equation}
\label{eq:lyrics_generator}
p(S) = \prod_{t=1}^{T} p(s_t | s_1, s_2, ..., s_{t-1}).
\end{equation}

\noindent The learned probability distribution is sampled one token at every time to generate the full sequence lyrics.

Lyrics to melody generator is modeled as a conditional distribution model over melody sequence on a lyrics sequence given by $p(m_1, m_2, .., m_{T} | s_1, s_2, ..., s_T)$ where $M = \{m_1, m_2, .., m_{T}\}$ is a melody sequence, which is modeled as a sequential encoder-decoder model. The encoder is an LSTM network taking the lyrics tokens generated by the lyrics generator model at each time step sequentially to produce the encoder hidden states $h_t$ at time $t$. The encoder also generates a variable context vector $C_i$ which encodes the lyrics for which most attention is paid during melody decoding. The decoder is another LSTM trained to predict the melody token at time $t$ given the hidden state. The hidden state of the decoder is computed recursively by
\begin{equation}
\label{eq:dec_hidden_state}
h_t = g(h_{t-1}, m_{t-1}, C_i).
\end{equation}
\noindent The next token distribution of the decoder can be  defined as
\begin{equation}
p(m_t | m_1, m_2, ...., m_{t-1}, C_i) = f(h_t, m_{t-1}, C_i).
\end{equation}
\noindent The model can be jointly trained to maximize the likelihood
\begin{equation}
\operatorname*{max}_\theta \frac{1}{N} \sum_{i=1}^{N} \log p_\theta(m_i | s_i),
\end{equation}
\noindent where $\theta$ is the model parameters and the pair $m_i, s_i$ is the input and output to the melody composer.

\subsection{Lyric to Vector (lyric2vec)}
\label{sec:feat_repre}

We train skip-gram models to obtain the dense representation of the input lyrics text. This enables us to learn high-quality vector representation of text tokens in an efficient way from large corpus of text data~\cite{mikolov2013efficient}. The objective of the skip-gram model is to predict the surrounding context tokens given a token at position $t$ in a piece of input text. We want the model to maximize the log probability of the surrounding tokens given by
\begin{equation}
\frac{1}{T} \sum_{t=1}^{T} \sum_{-c \le i \le c, i \neq 0} \log p({s_{t+i}} | {s_{t}}),
\end{equation}
\noindent where $c$ is the length of the context of the $t^{th}$ token $s_t$.








We train skip-gram models to obtain the embedding vectors of lyrics at syllable and word levels where $S = \{s_1, s_2, ..., s_n\}$ as syllable tokens and $W = \{w_1, w_2, w_3, ..., w_n\}$ as word tokens. During training, the learning rate with an initial value of 0.03 gradually decayed to 0.0007. We use $c = 7$ as context window length and negative sampling distribution parameter $\alpha$ is 0.75. We trained the models to obtain the syllable and word-level embedding vectors of dimensions $v = 50$.

The input one-hot vector representing the syllable or word is compressed to a low dimensional dense representation by linear units to predict the context tokens.
The following embedding representations are explored individually to train the lyrics prediction and melody generation model: i) syllable embeddings (SE) where each syllable $s_i$ is encoded with $v$ dimensional vector from syllable embedding model. ii) syllable and corresponding word embedding concatenation (SWC) where a syllable $s_i$ is concatenated with corresponding word embedding $w_i$ of $v$ dimension. iii) addition of syllable and word vector (ASW) where syllable vector is added element-wise to the word vector. iv) concatenated syllable, word and syllable projected word vector (CSWP). In CSWP, we project the syllable embedding vector $s_i$ onto the corresponding word vector $w_i$ by
\begin{equation}
proj_{w_i} s_i = \frac{s_i \cdot w_i}{|w_i|^2} w_i.
\end{equation}
\noindent Finally, we concatenate $s_i$, $w_i$ and $proj_{w_i} s_i$ to form an embedding vector for syllable $s_i$.

\subsection{Lyrics to Melody Net}
\label{sec:prop_lyrics_generator_melody_composer}

The proposed LTMN is a sequential encoder-decoder model to predict complete lyrics and generate the corresponding melody. Initially, LTMN takes seed lyrics as input and starts generating the lyrics one sentence at a time. The lyrics predictor takes the sequence of lyric tokens $S$ as input where each token is embedded into a dense vector by the skip-gram model. Each generated lyrics token is encoded by an RNN encoder to a hidden vector $h_t$ at each time step. The melody decoder uses the encoded hidden vector $h_t$ and the dynamic context vector $C_t$ to generate the melody hidden state at each time step. The independent decoder for each attribute receives the lyric context vector $C_t$ and decoder state to generate the pitch, duration, and rest attributes for each syllable. Each unit in the RNN is modeled with gated LSTM~\cite{cho2014learning}. 

The attention vector $c_t$ is used to align the input lyrics with the melody by computing alignment model~\cite{bahdanau2014neural}
\begin{equation}
c_t = \sum_{j=1}^{T} \alpha^{tj} h_t,
\end{equation}
\begin{equation}
\alpha^{tj} = \frac{exp(e^{tj})}{\sum_{k = 1}^{T} exp(e_{tk})},
\end{equation}
\begin{equation}
e^{tj} = V^{T}_a \tanh(W_a \tilde{h}_{t-1} + U_a h_t),
\end{equation}
\noindent where the parameters $V, W_a, U_a$ are the learnable weights.

The neural lyrics and melody composition model is trained end-to-end to minimize the total loss function of cross entropy.

\section{Experiments and Results}
\label{sec:experiments}

\subsection{Dataset}
The dataset used in our experiments was created in~\cite{yu2019conditional}. For our experiments, we parse the dataset into three different formats: a) The syllable level: This format is the lowest level that pairs together every note and the corresponding syllables and their attributes. b) The word level: This format regroups every note of a word and gives the attributes of every syllable that makes the word. c) The Sentence level: Similarly, this format put together every notes that forms a syllable (or in most case, a lyric line) and its corresponding attributes. Each syllable is represented with discrete attributes: a) The pitch of the note: We use MIDI note number as the representation of pitch, which can take any integer value between 0 and 127. b) The duration of the note: The duration of the note in number of staves. The exhaustive set of values it can take in our parsing is [0.125, 0.25, 0.5, 0.75, 1, 1.5, 2, 4, 8, 16, 32]. and c) Rest before the note: This value can take the same numerical values as the duration besides it can also be zero.

\subsection{Experiment setup}
The number of LSTM units used for all the models is 128. The melody attributes in the melody decoder model are treated as tokens and each attribute is represented with 128-dimensional word embedding.
The loss function is minimized by Adam optimizer with an initial learning rate of 0.0001 and linearly decayed every 10 epochs. The model is trained to minimize the cross entropy loss function with a batch size of 32. Our initial experiment of various temperature values $\tau$ shows that $\tau \in [0.5, 1.0]$ is the best range to generate the lyrics-melody pairs that follow dataset distribution, which is used in the following experiments.

\begin{table}[htbp]
\centering
\caption{Seed lyrics and the generated full length of lyrics.}
\label{table:seed_lyrics}
\includegraphics[width=0.45\textwidth]{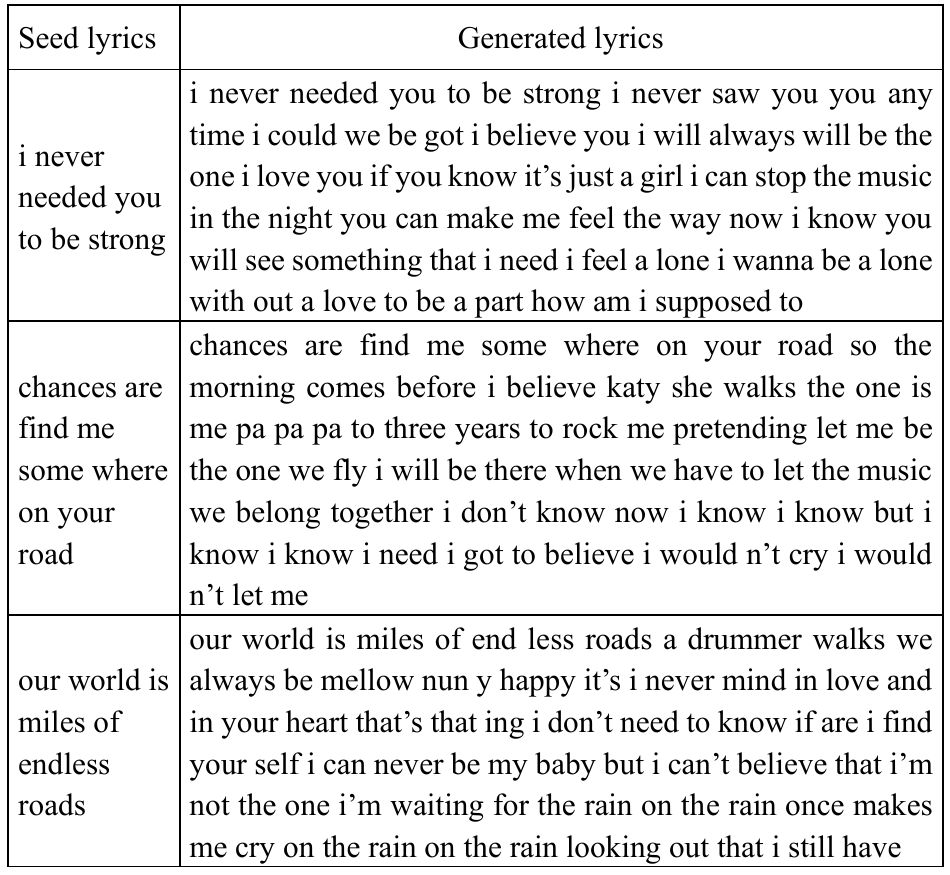}
\end{table}

\begin{figure}[htbp]
    \centering
    \subfloat[SE]{{\includegraphics[height=1.0cm, width=8.5cm]{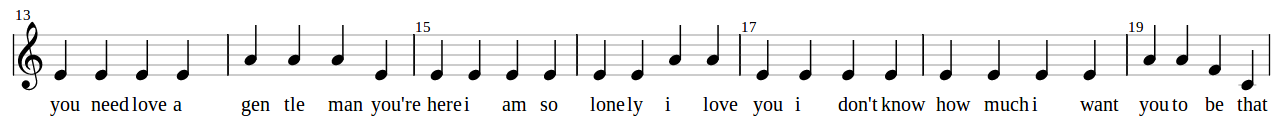}}}%
    \qquad
    \\
    \vspace{-4mm}
    \subfloat[SWC]{{\includegraphics[height=1.0cm, width=8.5cm]{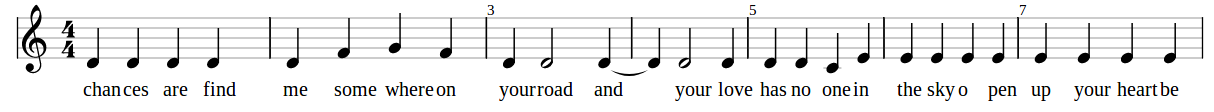}}}%
    \\
    \vspace{-4mm}
    \qquad
    \subfloat[ASW]{{\includegraphics[height=1.0cm, width=8.5cm]{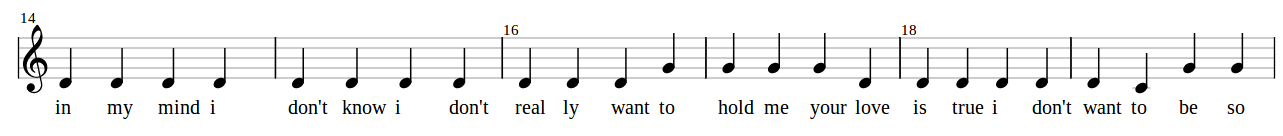}}}%
    \qquad
    \\
    \vspace{-4mm}
	\subfloat[CSWP]{{\includegraphics[height=1.0cm,
    width=8.6cm]{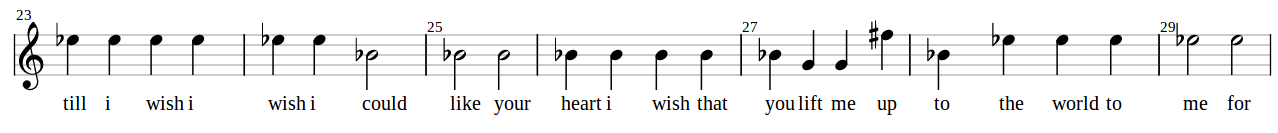}}}\\%

    \caption{Music sheet scores for generated lyrics-melody pairs for $\tau=0.8$ for SE, SWC, ASW and CSWP. 
    }%
    \label{fig:generated_lyrics_melody_pairs}%
\end{figure}

\subsection{Evaluation of Generation Quality}
\label{sec:translation quality}

The examples of seed lyrics and the generated full-length lyrics are shown in Table~\ref{table:seed_lyrics}. The generation results proved that our lyrics predictor can generate meaningful lyrics given the keywords as seed. 

\begin{figure}[htbp]
        \centering
        \includegraphics[width=0.45\textwidth, height=3.6cm]{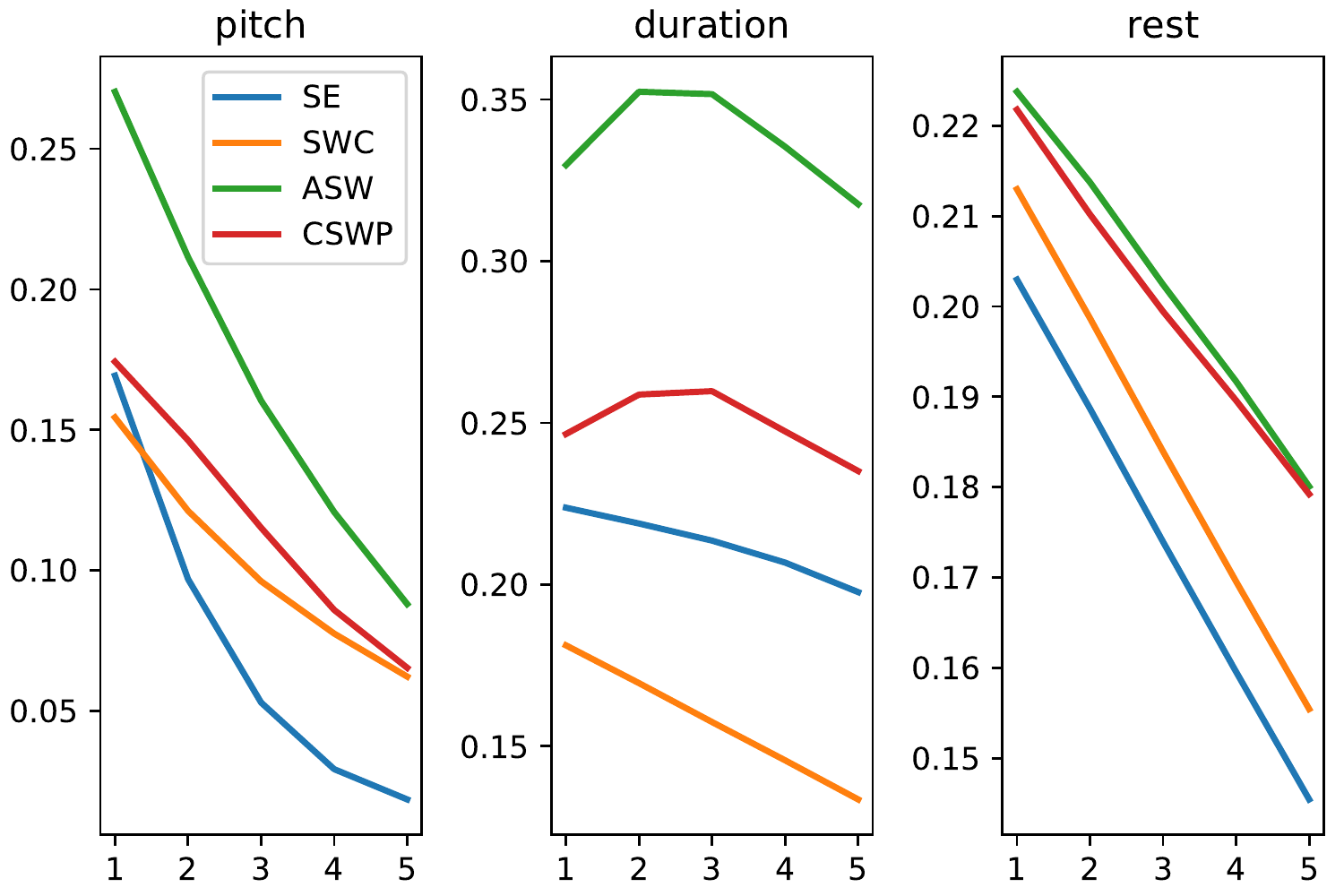}
        \caption{BLEU scores of pitch, duration and rest for SE, SWC, ASW and CSWP embedding representations.}
        \label{fig:bleu_score}
 \end{figure}

We evaluate our model with the BLEU score~\cite{papineni2002bleu} for predicted pitch, duration, and rest. Here, we compute the BLEU scores for the melodies generated from lyrics in the test set and the corresponding ground truth melodies. Higher BLEU score indicates better correlation between the generated and ground truth melodies. The results for each embedding representation are shown in Fig.~\ref{fig:bleu_score}. We can observe that ASW and CSWP lyrics embeddings in each music attribute generation outperform other representations. 
A few samples of sheet scores for the generated lyrics and melody pairs by the proposed LTMN is demonstrated in Fig.~\ref{fig:generated_lyrics_melody_pairs}.

\subsection{Subjective Evaluation}
Lyrics writing and melody composition are human creative processes, and it is very challenging to evaluate the quality of generated lyrics and melodies objectively and statistically. Hence, we adopt the subjective evaluation methods proposed in~\cite{yu2019conditional} and~\cite{lee2019icomposer}
to evaluate lyrics and melodies generated by LTMN. We ask the following questions to participants during subjective evaluation:i) how meaningful are the predicted lyrics? and ii) how well do the melodies fit the lyrics?
Participants grade the given samples on a five-point scale. We randomize 5 generated full-length lyrics and lyrics-melody pairs from the proposed LTMN, baseline, and ground truth.

We follow~\cite{yu2019conditional} to create baseline lyrics and melodies. The attributes of the baseline melodies i.e., pitch, duration, and rest are sampled from the respective data distributions. The baseline lyrics are sampled from the most frequent syllable lyrics vocabulary. We separately conduct the subjective evaluation for generated lyrics and lyrics-melody pairs. 

\begin{figure}[htbp]
        \centering
		\includegraphics[width=0.45\textwidth]{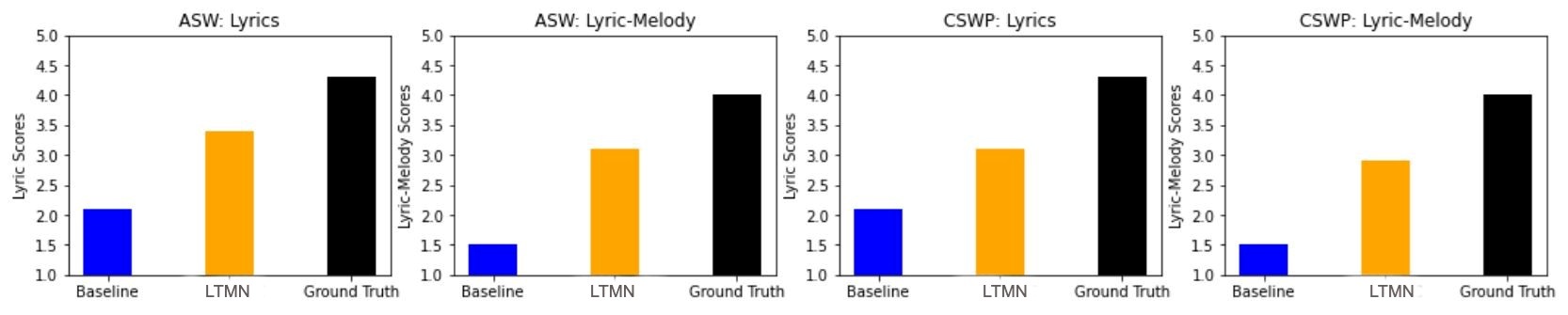}
        \caption{Subjective evaluation results for lyrics and lyric-melody pairs generated from ASW and CSWP of the proposed LTMN.}
        \label{fig:subjective_score}
 \end{figure}
11 adults who have at least basic knowledge about lyrics, melodies, and music composition participated in the subjective evaluation. The subjective evaluation results for ASW and CSWP for $\tau = 0.6$ are shown in Fig.~\ref{fig:subjective_score}. We can observe that LTMN is close to human compositions. The baseline performs worse than the other methods without surprise. We also see that the ASW performs better than CSWP which is consistent with the objective evaluation in Section \ref{sec:experiments}. From the subjective evaluation measures, we can find that there is a gap between the human compositions and generated lyrics and melodies by the proposed model, which indicates that it is potential to explore injecting prior musical knowledge to improve the current model.

\section{Summary}
\label{sec:sum_conc}

 We proposed deep attention-based alignment network for training LTMN, with the aim of enabling people to discover the proper lyrics and generate meaningful melodies. We trained lyric2vec models on a large set of lyrics-melody parallel dataset parsed at syllable, word and sentence levels to extract the lyric embeddings. The proposed LTMN is an encoder-decoder sequential recurrent neural network model consisting of lyrics predictor and melody generator for each note attribute, trained end-to-end to learn the correlation between the lyrics and melody pairs with attention mechanism. In addition, experiments proved that lyrics embedding extracted from ASW performs better than CSWP, SE, and SWC. The qualitative and quantitative evaluation showed that the proposed method is capable of generating proper lyrics and meaningful melodies.

\bibliographystyle{IEEEtran}
\bibliography{mybib}

\begin{thebibliography}{10}
\providecommand{\url}[1]{#1}
\csname url@samestyle\endcsname
\providecommand{\newblock}{\relax}
\providecommand{\bibinfo}[2]{#2}
\providecommand{\BIBentrySTDinterwordspacing}{\spaceskip=0pt\relax}
\providecommand{\BIBentryALTinterwordstretchfactor}{4}
\providecommand{\BIBentryALTinterwordspacing}{\spaceskip=\fontdimen2\font plus
\BIBentryALTinterwordstretchfactor\fontdimen3\font minus
  \fontdimen4\font\relax}
\providecommand{\BIBforeignlanguage}[2]{{%
\expandafter\ifx\csname l@#1\endcsname\relax
\typeout{** WARNING: IEEEtran.bst: No hyphenation pattern has been}%
\typeout{** loaded for the language `#1'. Using the pattern for}%
\typeout{** the default language instead.}%
\else
\language=\csname l@#1\endcsname
\fi
#2}}
\providecommand{\BIBdecl}{\relax}
\BIBdecl

\bibitem{pachet2011markov}
F.~Pachet and P.~Roy, ``Markov constraints: Steerable generation of {Markov}
  sequences,'' \emph{Constraints}, vol.~16, no.~2, pp. 148--172, 2011.

\bibitem{pachet2017sampling}
F.~Pachet, A.~Papadopoulos, and P.~Roy, ``Sampling variations of sequences for
  structured music generation.'' in \emph{ISMIR}, 2017, pp. 167--173.

\bibitem{waite2016generating}
E.~Waite, ``Generating long-term structure in songs and stories,''
  \emph{Magenta Bolg}, 2016.

\bibitem{chu2016song}
H.~Chu, R.~Urtasun, and S.~Fidler, ``Song from {PI}: A musically plausible
  network for pop music generation,'' \emph{arXiv preprint arXiv:1611.03477},
  2016.

\bibitem{mogren2016c}
O.~Mogren, ``{C-RNN-GAN}: Continuous recurrent neural networks with adversarial
  training,'' \emph{arXiv preprint arXiv:1611.09904}, 2016.

\bibitem{dong2018musegan}
H.-W. Dong, W.-Y. Hsiao, L.-C. Yang, and Y.-H. Yang, ``Musegan: Multi-track
  sequential generative adversarial networks for symbolic music generation and
  accompaniment,'' in \emph{AAAI}, 2018.

\bibitem{yu2019deep}
Y.~Yu, S.~Tang, F.~Raposo, and L.~Chen, ``Deep cross-modal correlation learning
  for audio and lyrics in music retrieval,'' \emph{ACM Transactions on
  Multimedia Computing, Communications, and Applications}, vol.~15, no.~1,
  p.~20, 2019.

\bibitem{ackerman2017algorithmic}
M.~Ackerman and D.~Loker, ``Algorithmic songwriting with alysia,'' in
  \emph{EvoMUSART}, 2017, pp. 1--16.

\bibitem{bao2018neural}
H.~Bao, S.~Huang, F.~Wei, L.~Cui, Y.~Wu, C.~Tan, S.~Piao, and M.~Zhou, ``Neural
  melody composition from lyrics,'' \emph{arXiv:1809.04318}, 2018.

\bibitem{yu2019conditional}
Y.~Yu, A.~Srivastava, and S.~Canales, ``Conditional {LSTM-GAN} for melody
  generation from lyrics,'' \emph{arXiv:1908.05551}, 2019.

\bibitem{yu2020lyrics}
Y.~Yu, F.~Harsco{\"e}t, S.~Canales, G.~Reddy, S.~Tang, and J.~Jiang,
  ``Lyrics-conditioned neural melody generation,'' in \emph{MMM}, 2020, pp.
  709--714.

\bibitem{srivastava_melody_2022}
A.~Srivastava, W.~Duan, R.~R. Shah, J.~Wu, S.~Tang, W.~Li, and Y.~Yu, ``Melody
  {Generation} from {Lyrics} {Using} {Three} {Branch} {Conditional}
  {LSTM}-{GAN},'' in \emph{MMM}, 2022, pp. 569--581.

\bibitem{ShengST2021}
Z.~Sheng, K.~Song, X.~Tan, Y.~Ren, W.~Ye, S.~Zhang, and T.~Qin, ``{SongMASS}:
  Automatic song writing with pre-training and alignment constraint,'' in
  \emph{AAAI}, 2021, pp. 13\,798--13\,805.

\bibitem{duan_interpretable_2022}
W.~Duan, Z.~Zhang, Y.~Yu, and K.~Oyama, ``Interpretable {{Melody Generation}}
  from {{Lyrics}} with {{Discrete-Valued Adversarial Training}},'' in \emph{ACM
  MM}, Oct. 2022, pp. 6973--6975.

\bibitem{yu_conditional_2022}
Y.~Yu, Z.~Zhang, W.~Duan, A.~Srivastava, R.~Shah, and Y.~Ren, ``Conditional
  hybrid {GAN} for melody generation from lyrics,'' \emph{Neural Computing and
  Applications}, Oct. 2022.

\bibitem{mikolov2013efficient}
T.~Mikolov, K.~Chen, G.~Corrado, and J.~Dean, ``Efficient estimation of word
  representations in vector space,'' \emph{arXiv preprint arXiv:1301.3781},
  2013.

\bibitem{cho2014learning}
K.~Cho, B.~Van~Merri{\"e}nboer, C.~Gulcehre, D.~Bahdanau, F.~Bougares,
  H.~Schwenk, and Y.~Bengio, ``Learning phrase representations using {RNN}
  encoder-decoder for statistical machine translation,'' \emph{arXiv preprint
  arXiv:1406.1078}, 2014.

\bibitem{bahdanau2014neural}
D.~Bahdanau, K.~Cho, and Y.~Bengio, ``Neural machine translation by jointly
  learning to align and translate,'' \emph{arXiv preprint arXiv:1409.0473},
  2014.

\bibitem{papineni2002bleu}
K.~Papineni, S.~Roukos, T.~Ward, and W.-J. Zhu, ``Bleu: a method for automatic
  evaluation of machine translation,'' in \emph{ACL}, 2002, pp. 311--318.

\bibitem{lee2019icomposer}
H.-P. Lee, J.-S. Fang, and W.-Y. Ma, ``icomposer: An automatic songwriting
  system for chinese popular music,'' in \emph{NAACL}, 2019, pp. 84--88.

\end{thebibliography}

\end{document}